\documentclass[english]{IEEEtran}
\usepackage[T1]{fontenc}
\usepackage[latin9]{inputenc}
\usepackage{xcolor}
\usepackage{float}
\usepackage{units}
\usepackage{mathrsfs}
\usepackage{amsmath}
\usepackage{amssymb}
\usepackage{graphicx}

\makeatletter

\providecommand{\tabularnewline}{\\}
\floatstyle{ruled}
\newfloat{algorithm}{tbp}{loa}
\providecommand{\algorithmname}{Algorithm}
\floatname{algorithm}{\protect\algorithmname}

\usepackage{eqnarray}
\usepackage{mathrsfs}
\usepackage{epsfig}
\usepackage[english]{babel}
\usepackage{subfigure}
\usepackage{epstopdf}
\usepackage{import}
\usepackage{color}
\usepackage{colortbl}\usepackage{cite}
\usepackage{algorithm}

\usepackage{bbm}
\usepackage{cases}
\usepackage{array}
\usepackage[english]{babel}
\usepackage{times}
\usepackage{color}
\usepackage{amsfonts}
\usepackage{psfrag}
\usepackage{fancyhdr}
 \usepackage{algorithmic}
\allowdisplaybreaks

\usepackage{dsfont}
\usepackage{babel}

\makeatother

\usepackage{babel}
\begin{document}
\title{\title{ \fontsize{19}{20}\selectfont  Ultra-Reliable and Low-Latency Vehicular Communication: An Active Learning Approach } }
\author{\IEEEauthorblockN{Mohamed~K.~Abdel-Aziz,~\IEEEmembership{Student Member,~IEEE},
Sumudu~Samarakoon,~\IEEEmembership{Member,~IEEE}, Mehdi~Bennis,~\IEEEmembership{Senior Member,~IEEE},
and Walid~Saad,~\IEEEmembership{Fellow,~IEEE}}\thanks{This work was supported in part by the Academy of Finland project
CARMA, and 6Genesis Flagship (grant no. 318927), in part by the INFOTECH
project NOOR, in part by the Office of Naval Research (ONR) under
MURI Grant N00014-19-1-2621, and in part by the Kvantum Institute
strategic project SAFARI.}\thanks{M. K. Abdel-Aziz and S. Samarakoon are with the Centre for Wireless
Communications, University of Oulu, 90014 Oulu, Finland (e-mails:
mohamed.abdelaziz@oulu.fi; sumudu.samarakoon@oulu.fi).}\thanks{M. Bennis is with the Centre for Wireless Communications, University
of Oulu, 90014 Oulu, Finland, and also with the Department of Computer
Engineering, Kyung Hee University, Yongin 446-701, South Korea (e-mail:
mehdi.bennis@oulu.fi).}\thanks{W. Saad is with the Department of Electrical and Computer Engineering,
Virginia Tech, Blacksburg, VA 24061, USA (e-mail: walids@vt.edu).}\vspace{-1cm}
}
\maketitle
\begin{abstract}
In this letter, an age of information (AoI)-aware transmission power
and resource block (RB) allocation technique for vehicular communication
networks is proposed. Due to the highly dynamic nature of vehicular
networks, gaining a prior knowledge about the network dynamics, i.e.,
wireless channels and interference, in order to allocate resources,
is challenging. Therefore, to effectively allocate power and RBs,
the proposed approach allows the network to actively learn its dynamics
by balancing a tradeoff between minimizing the probability that the
vehicles\textquoteright{} AoI exceeds a predefined threshold and maximizing
the knowledge about the network dynamics. In this regard, using a
Gaussian process regression (GPR) approach, an online decentralized
strategy is proposed to actively learn the network dynamics, estimate
the vehicles\textquoteright{} future AoI, and proactively allocate
resources. Simulation results show a significant improvement in terms
of AoI violation probability, compared to several baselines, with
a reduction of at least $50\%$.\vspace{-0.3cm}
\end{abstract}

\begin{IEEEkeywords}
Gaussian process regression (GPR), ultra-reliable low-latency communication
(URLLC), age of information (AoI), V2X.\vspace{-0.5cm}
\end{IEEEkeywords}

\section{Introduction}

Vehicle-to-vehicle (V2V) communication is an important application
in 5G and beyond networks. Time-critical V2V safety applications require
ultra-reliable low-latency communication (URLLC), in which the freshness
of vehicles\textquoteright{} status update is crucial. A relevant
metric to quantify information freshness is the age of information
(AoI). AoI is defined as the time elapsed since the generation of
the last received status update \cite{AoI_First,OurWork}. Providing
reliability guarantees in terms of AoI is essential for such applications.
In fact, if future AoI can be reliably estimated, proactive transmission
power and resource block (RB) allocation can be carried out to ensure
that the reliability, defined as the probability of the future AoI
exceeding a threshold, is minimized. However, for an accurate estimation
of future AoI, knowledge of the network dynamics, i.e., wireless channels
and interference, is required. Recently, several works have studied
the optimization of AoI in vehicular networks \cite{AoIV2X_2,AoIV2X_3,AoIV2X_4}.
In \cite{AoIV2X_2} and \cite{AoIV2X_3}, the network dynamics are
assumed to be known and in \cite{AoIV2X_4}, they are estimated in
a centralized manner without any consideration of future AoI. Yet,
in a URLLC setting \cite{PetarURLLC,6G_WalidMehdi}, reliably learning
and estimating the network dynamics with minimum communication overhead
is desirable. Due to the highly dynamic nature of vehicular networks,
gaining knowledge a priori about the network dynamics is challenging.
In this regard, a viable solution is to develop an online decentralized
strategy to actively learn\footnote{Active learning is a paradigm in which the learning process is directed
to provide the next-best data input, e.g., resource allocation in
our case, that optimizes an objective according to the historical
data.} the network dynamics and allocate resources accordingly to ensure
reliability \cite{TansuetShames}. One possible approach is based
on model predictive control (MPC) \cite{MPC}, which allows to control
dynamic systems while ensuring the optimality of the system performance.
However, MPC requires an accurate mathematical model of the dynamic
system a priori, which is unrealistic in vehicular networks. One alternative
is based on a Gaussian process regression (GPR) approach, leveraging
its modeling flexibility and robustness to overfitting \cite{TansuetShames}.
Unlike Gaussian mixture model (GMM), which is a parametric model and
suffers from overfitting, GPR is a non-parametric Bayesian learning
approach which means that it is not limited by a functional form \cite{Rasmussen2005}.

\begin{figure}
\begin{centering}
\includegraphics[width=0.36\textwidth]{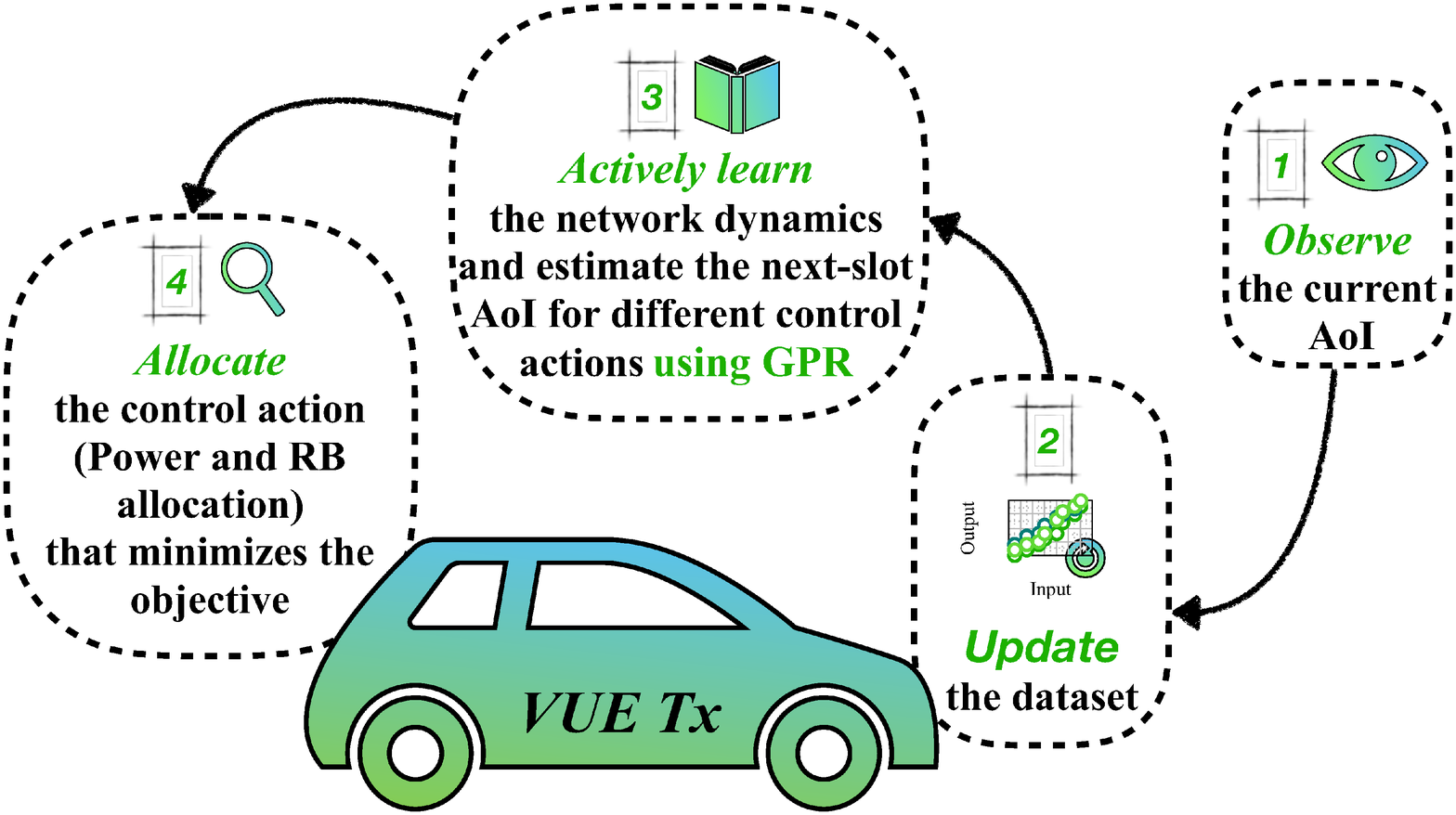}\vspace{-0.3cm}
\par\end{centering}
\centering{}\caption{Illustration of the proposed solution.\label{fig:Per-VUE-power}}
\vspace{-0.5cm}
\end{figure}

The main contribution of this paper is to develop a novel approach,
based on GPR, for enabling URLLC among vehicles by actively learning
the system dynamics, estimating the vehicles' future AoI, and proactively
allocating transmission power and RBs, in an online decentralized
manner, as illustrated in Fig.~\ref{fig:Per-VUE-power}. The main
objective is to trade off between maximizing reliability by minimizing
the probability that the AoI exceeds a predefined threshold and maximizing
the knowledge gain about the local network dynamics. Simulation results
show a significant gain in terms of maximizing reliability, compared
to other baselines, with more than a 2-fold improvement. Moreover,
our numerical results show that accurately estimating the future AoI
improves the overall performance in terms of minimizing the AoI violation
probability.\vspace{-0.7cm}

\section{System Model\label{sec:System-Model}}

Consider a V2V communication network consisting of a set $\mathcal{K}$
of $K$ VUE transmitter-receiver pairs under the coverage of a single
roadside unit (RSU), assuming a Manhattan mobility model\textcolor{blue}{}\footnote{A transmitter periodically shares its status with a receiver in a
vehicular rear-end collision aviodance scenario.}. The VUE pairs share a set $\mathcal{N}$ of $N$ orthogonal RBs
with equal bandwidth $W$. The communication timeline is slotted with
a slot duration of $\tau$, with $t$ being the slot index. For each
VUE pair $k$, $\boldsymbol{P}_{k}(t)=[P_{k}^{n}(t)]_{n=1}^{N}\in\mathcal{P}$
is an $N$-dimensional power allocation vector over different RBs,
with $\mathcal{P}$ being the set of all feasible power allocations\textcolor{blue}{{}
}and $P_{k}^{n}(t)\in\{0,\frac{p}{L},\frac{2p}{L},\ldots,p\}$, where
$p$ is the maximum allowed transmission power per RB and $L+1$ is
the number of available power levels\footnote{Note that $P_{k}^{n}(t)=0$ implies that RB $n$ is not allocated
to VUE pair $k$.}. Moreover, the allocated transmission power is subject to $\sum_{n\in\mathcal{N}}P_{k}^{n}(t)\leq P_{\text{max}}$,
where $P_{\text{max}}$ is the total power budget, which is assumed
equal for all VUE pairs. Furthermore, each VUE transmitter $k$ has
a queue buffer to store the data destined to its receiver whose queue
dynamics are given by: $Q_{k}(t+1)=\max\left(Q_{k}(t)-R_{k}(t),0\right)+A,$
where $Q_{k}(t)$ is the queue length of VUE pair $k$ at time slot
$t$, $A$ is the periodic packet arrival rate per slot, and $R_{k}(t)$
is the transmission rate of each VUE pair $k$ at time slot $t$ (in
packets per slot), which is given by, \vspace{-0.3cm}
\begin{equation}
R_{k}(t)=\frac{\tau}{Z}{\textstyle \sum\limits _{n\in\mathcal{N}}}W\log_{2}\left(1+\frac{P_{k}^{n}(t)h_{kk}^{n}(t)}{N_{\text{0}}W+I_{k}^{n}(t)}\right),\label{eq:TransmissionRate}
\end{equation}
where $Z$ is the packet length in bits, and $N_{\text{0}}$ is the
power spectral density of the additive white Gaussian noise. Here,
$I_{k}^{n}(t)=\sum_{k'\in\mathcal{K}\setminus\{k\}}P_{k'}^{n}(t)h_{k'k}^{n}(t)$
is the aggregate interference from other VUEs at the receiver of VUE
pair $k$ over RB $n$, and $h_{kk}^{n}(t)$ is the channel gain from
the transmitter of VUE pair $k$ to its receiver over RB $n$. We
use the realistic V2V channel model of \cite{Path_loss_model} in
which, depending on the location of the VUE transmitter and receiver,
the channel model is categorized into three types: Line-of-sight,
weak-line-of-sight, and non-line-of-sight.

Real-time status updates are critical for V2V safety applications,
whose freshness can be characterized using the AoI metric \cite{AoI_First},
defined as: $\triangle_{k}(t)\triangleq\tau t-\gamma_{k}(t)$, where
$\triangle_{k}(t)$ denotes the AoI of VUE pair $k$ at the beginning
of time slot $t$ and $\gamma_{k}(t)$ is the generation instant of
the last status update received by VUE receiver $k$ at/or just before
the beginning of time slot $t$. Note that, the index of the last
received status update depends on $R_{k}(t-1)$, which, according
to (\ref{eq:TransmissionRate}), depends on the channel gain, interference,
and allocated power. Therefore, $\triangle_{k}(t+1)$ can be expressed
as: \vspace{-0.1cm}
\begin{equation}
\triangle_{k}(t+1)=f_{k}\left(\triangle_{k}(t),\boldsymbol{P}_{k}(t)\right),~\forall k\in\mathcal{K},\label{eq:GPR1}
\end{equation}
where $f_{k}(\cdot)$ is a nonlinear dynamic system for VUE pair $k$
which represents the local network dynamics, i.e. wireless channels
and interference. In this dynamic system, $\triangle_{k}(t)\in\mathbb{R}^{+}$
and $\boldsymbol{P}_{k}(t)\in\mathcal{P}$ represent the state of
the system and the control action at time slot $t$, respectively.
$f_{k}:\mathbb{R}^{+}\times\mathcal{P}\rightarrow\mathbb{R}^{+}$
maps the control actions to the state of the system \cite{TansuetShames}.
Without any prior knowledge about the channel $h_{kk}^{n}(t)$ and
interference $I_{k}^{n}(t)$, and due to the highly non-stationary
nature of a vehicular network, the relationship $f_{k}(\cdot)$ is
unknown a priori for every VUE pair $k$. Thus, $f_{k}(\cdot)$ needs
to be reliably learned online from historical data. Jointly controlling
and actively learning $f_{k}(\cdot)$ is crucial to reliably estimate
the future AoI and allocate resources.

Estimating $f_{k}(\cdot)$ using a set of historical data points is
a regression problem. For our model, GPR is used as the regression
method, which is a class of Bayesian non-parametric machine learning
models. GPR does not involve any black-box operations, and has a promising
potential in improving prediction accuracy. Moreover, Gaussian processes
(GPs) provide an elegant mathematical method for regression accompanied
with the full predictive distribution, which is important for establishing
confidence intervals \cite{GPSignalProcessing,TansuetShames}. In
this view, we can rewrite (\ref{eq:GPR1}), for notational simplicity,
as $y_{i}=f_{k}\left(\boldsymbol{x}_{i}\right),$ where $\boldsymbol{x}_{i}=[\triangle_{k}(i),\boldsymbol{P}_{k}(i)]$
represents the $i^{\text{th}}$ input and $y_{i}=\triangle_{k}(i+1)$
represents the corresponding scalar output. GPR is a nonlinear regressor
that expresses the input-output relation by assuming that $f_{k}\left(\cdot\right)$
a priori follows a GP\footnote{Note that this assumption does not mean that the underlying process
is precisely Gaussian, but GP can still be used as a maximum entropy
process, for a given covariance function.} \cite{GPSignalProcessing}. For any finite dataset $\mathcal{D}_{k}=\left\{ \boldsymbol{x}_{i},y_{i}\right\} _{i=1}^{M}$,
where $M$ is the dataset size, a GP becomes a multidimensional Gaussian
defined by its mean (zero in our case, for simplicity) and covariance
matrix, $\boldsymbol{C}\triangleq\left[c\left(\boldsymbol{x}_{i},\boldsymbol{x}_{j}\right)\right]_{ij},\forall\boldsymbol{x}_{i},\boldsymbol{x}_{j}\in\mathcal{D}_{k}$.
For a general input $\boldsymbol{x}_{*}$ and for a given $\mathcal{D}_{k}$,
GPR provides a full statistical description of $y_{*}$, namely $\textrm{Pr}\left\{ y_{*}\mid\boldsymbol{x}_{*},\mathcal{D}_{k}\right\} $,
which can be computed using the standard tools of Bayesian statistics
leading to $y_{*}\mid_{\boldsymbol{x}_{*},\mathcal{D}_{k}}\sim\mathscr{N}\left(\mu_{y_{*}},\sigma_{y_{*}}^{2}\right)$,
where $\mathscr{N}\left(\mu_{y_{*}},\sigma_{y_{*}}^{2}\right)$ is
the normal distribution with mean $\mu_{y_{*}}$ and variance $\sigma_{y_{*}}^{2}$.
Moreover, \vspace{-0.4cm}
\begin{align}
\mu_{y_{*}} & =\boldsymbol{c}_{*}^{\text{T}}\boldsymbol{C}^{-1}\boldsymbol{y},\label{eq:mean}\\
\sigma_{y_{*}}^{2} & =c\left(\boldsymbol{x}_{*},\boldsymbol{x}_{*}\right)-\boldsymbol{c}_{*}^{\text{T}}\boldsymbol{C}^{-1}\boldsymbol{c}_{*},\label{eq:variance}
\end{align}
with $\boldsymbol{c}_{*}\triangleq\left[c\left(\boldsymbol{x}_{*},\boldsymbol{x}_{1}\right)\cdots c\left(\boldsymbol{x}_{*},\boldsymbol{x}_{M}\right)\right]^{\text{T}}$
and $\boldsymbol{y}=[y_{1}\cdots y_{M}]^{\text{T}}$ \cite{GPSignalProcessing}.
Here, the mean given in (\ref{eq:mean}) is the estimate of $y_{*}$.
Note that the GPR computation complexity grows cubically with the
dataset size $M$ \cite{Rasmussen2005}. However, due to the dynamic
nature of vehicular networks, fixing the maximum size of the dataset
by discarding old samples (i.e. history), can help tackle the computational
complexity without affecting the GPR estimation performance.

A GP is completely defined by its covariance function $c\left(\boldsymbol{x}_{i},\boldsymbol{x}_{j}\right)$.
To accurately estimate future measures and their distributions, a
covariance function that fits with the nature of the system has to
be selected. In this regard, a Matérn class covariance function \cite{Stein1999}
has been selected for learning $f_{k}(\cdot)$:
\[
c\left(\boldsymbol{x}_{i},\boldsymbol{x}_{j}\right)=h^{2}\frac{2^{1-\nu}}{\varGamma(\nu)}\left(2\sqrt{\nu}\frac{\left|\boldsymbol{x}_{i}-\boldsymbol{x}_{j}\right|}{\lambda}\right)^{\nu}\mathbb{B}\left(2\sqrt{\nu}\frac{\left|\boldsymbol{x}_{i}-\boldsymbol{x}_{j}\right|}{\lambda}\right),
\]
where $\varGamma(\cdot)$ is the standard Gamma function and $\mathbb{B}(\cdot)$
is the modified Bessel function of second order\textcolor{blue}{.
}The Matérn class covariance function includes both the exponential
autocorrelation function when $\nu=0.5$ and the Gaussian autocorrelation
function as a limiting case when $\nu\to\infty$. Therefore, the Matérn
class offers us a flexibility to strike a balance between these two
extremes, thus it is well suited for our application. The variables
$h,\lambda\text{ and }\nu$ are the hyperparameters of the covariance
function. The hyperparameters determine the shape of the covariance
function which need to be tuned to fit the observed dataset $\mathcal{D}_{k}$.
This is done by maximizing the marginal likelihood of the GP \cite{Rasmussen2005}.

Actively learning $f_{k}(\cdot)$, by choosing the next-best input
$\boldsymbol{x}_{*}$ that optimizes an objective, would be beneficial
in vehicular networks due to its high non-stationary nature. The choice
of this objective and the problem formulation is discussed next.\vspace{-0.1cm}

\section{AoI-Aware Power and RB Allocation\label{sec:Problem-Formulation}}

Each VUE pair has two main objectives. First, it seeks to improve
the reliability of the status updates and second, it needs to enhance
its knowledge about the system dynamics. Improving the status updates
reliability is captured by choosing the resources allocation that
minimizes $\textrm{Pr}\left\{ \hat{\triangle}_{k}(t+1)>d\right\} $.
Here, the next-slot AoI is estimated using GPR and thus, $\hat{\triangle}_{k}(t+1)$
is a Gaussian distributed random variable with mean and variance given
by (\ref{eq:mean}) and (\ref{eq:variance}), respectively. Moreover,
enhancing the knowledge about $f_{k}(\cdot)$ is tantamount to choosing
the allocation that maximizes the estimated AoI variance, $\sigma_{\hat{\triangle}_{k}(t+1)}^{2}$
\cite[Corollary 5.4]{TansuetShames}. Formally, actively learning
$f_{k}(\cdot)$ by optimizing the tradeoff between improving status
update reliability (exploitation) and enhancing the system dynamics
knowledge (exploration) of VUE pair $k$ at time $t$ can be posed
as follows: 
\begin{align}
\min_{{\color{black}{\color{black}{\color{black}{\color{blue}{\color{black}{\color{brown}{\color{black}\boldsymbol{P}_{k}(t)\in\mathcal{P}}}}}}}}} & ~~\alpha_{c}\textrm{Pr}\left\{ \hat{\triangle}_{k}(t+1)>d\right\} -\alpha_{i}\sigma_{\hat{\triangle}_{k}(t+1)}^{2}\label{eq:Objective}\\
\text{subject to} & ~~\eqref{eq:GPR1},\nonumber 
\end{align}
where $\alpha_{c}$ and $\alpha_{i}$ are non-negative weighting factors
that capture the exploitation-exploration tradeoff. A control action
\textcolor{blue}{${\normalcolor {\normalcolor \boldsymbol{P}_{k}(t)\in\mathcal{P}}}$}
that enhances the knowledge about the unknown system $f_{k}(.)$ may
not necessarily be the one that improves the reliability $\textrm{Pr}\left\{ \hat{\triangle}_{k}(t+1)>d\right\} $.
Thus, the choice of $\alpha_{c}$ and $\alpha_{i}$ affect the system
performance, as will be shown in Section \ref{sec:Simulation-Results}.
Note that, since $\hat{\triangle}_{k}(t+1)$ follows a Gaussian distribution,
then $\textrm{Pr}\left\{ \hat{\triangle}_{k}(t+1)>d\right\} =\frac{1}{2}\text{erfc}\left(\frac{d-\mu_{\hat{\triangle}_{k}\left(t+1\right)}}{\sqrt{2\sigma_{\hat{\triangle}_{k}(t+1)}^{2}}}\right)$.

Actively learning $f_{k}(\cdot),\forall k\in\mathcal{K}$ at a central
entity, i.e., RSU,\textbf{ }incurs huge communication overhead due
to exchanging the AoI of each VUE pair and the resources allocation
decisions between the RSU and all VUE pairs at every time slot $t$.
Thus, a decentralized AoI estimation and resources allocation approach
is proposed\textcolor{blue}{}\footnote{A VUE transmitter only requires the state (AoI) of its receiver.}.
For a given VUE pair $k$, Algorithm \ref{alg:Algorithm-for-solvingP2},
which is given from \cite{TansuetShames}, provides the steps to obtain
the solution of (\ref{eq:Objective}).\vspace{-0.3cm}
 
\begin{algorithm}[t]
\begin{algorithmic}[1]\footnotesize

\STATE  \textbf{Input: }set of available transmission power actions
$\mathcal{P}$, $\alpha_{c}$, $\alpha_{i}$, $M$.

\STATE  \textbf{Initialization: }$\triangle_{k}(0)=0$, $\mathcal{D}_{k}=\emptyset$,
select $\boldsymbol{P}_{k}(0)\in\mathcal{P}$ randomly.

\STATE    \textbf{for} $t=1,2,\ldots$ \textbf{do}.

 \begin{ALC@g} \STATE  Observe $\triangle_{k}(t)$.

\STATE  Augment the dataset $\mathcal{D}_{k}=\mathcal{D}_{k}\cup\Bigl(\bigl[\triangle_{k}(t-1),\boldsymbol{P}_{k}(t-1)\bigr],\triangle_{k}(t)\Bigr)$.

\STATE \textbf{if $\left|\mathcal{D}_{k}\right|>M$}

 \begin{ALC@g} \STATE  Remove the oldest data point from $\mathcal{D}_{k}$.

 \end{ALC@g}\STATE  end \textbf{if}

\STATE  Calculate the mean and variance of $\hat{\triangle}_{k}(t+1)\,\forall\boldsymbol{P}_{k}(t)\in\mathcal{P}$
using (\ref{eq:mean}) and (\ref{eq:variance}).

\STATE  Evaluate $\textrm{Pr}\left\{ \hat{\triangle}_{k}(t+1)>d\right\} \,\text{for all }\boldsymbol{P}_{k}(t)\in\mathcal{P}$.

\STATE  Choose $\boldsymbol{P}_{k}^{*}(t)$ that minimizes (\ref{eq:Objective}).

 \end{ALC@g}\STATE  end \textbf{for}

\end{algorithmic}

\caption{\label{alg:Algorithm-for-solvingP2} Per VUE power and RB allocation
using GPR}
\end{algorithm}

\section{Simulation Results\label{sec:Simulation-Results}}

\begin{table}
\vspace{-0.3cm}
\caption{\label{tab:Simulation-Parameters}Simulation Parameters}
\vspace{-0.2cm}

\centering{}%
\begin{tabular}{|c|c|c|c|c|c|}
\hline 
Para. & Value & Para. & Value & Para. & Value\tabularnewline
\hline 
$K$ & 20 & $N$ & 20 & $\tau$ & 3 ms\tabularnewline
$W$ & 180 kHz & $N_{\text{0}}$ & -174 dBm/Hz & $p$ & 10 dBm\tabularnewline
$P_{\text{max}}$ & 17 dBm & $L$ & 1 & arrival rate & $2.5\,\text{Mbps}$\tabularnewline
$Z$ & 500 Byte & $\alpha_{c}$ & 1 & $\alpha_{i}$ & 100\tabularnewline
$d$ & 10 ms & $M$ & 200 &  & \tabularnewline
\hline 
\end{tabular}\vspace{-0.5cm}
\end{table}
We assume a $250\times250\,\text{m}^{2}$ area of a Manhattan mobility
model as in \cite{ChenEVT2018}. We set the average vehicle speed
to $60\,\text{km/h}$, and we choose a time-varying distance between
each transmitter-receiver pair while maintaining an average distance
of $15\,\text{m}$ \cite{ETSIInterVehicleDistance}\textcolor{blue}{,
}with a simulation duration of $T=5000\text{\,time slots}$\textcolor{blue}{.
}Table \ref{tab:Simulation-Parameters} lists all main simulation
parameters. The GPML toolbox has been used to implement GPR \cite{rasmussen2010gaussian}.
A performance comparison is carried out between different schemes:
\textbf{i) Baseline 1}: each VUE pair allocates its transmission power
randomly without learning the system dynamics, i.e. wireless channel
and interference, \textbf{ii) Baseline 2 (GPR-based Allocation with
no exploration)}: a greedy approach of our proposed scheme where each
VUE pair actively learns its system dynamics $f_{k}$ using GPR and
subsequently solves (\ref{eq:Objective}) with $\alpha_{i}=0$, and
\textbf{iii) Proposed}: GPR-based power and RB allocation with $\alpha_{i}=100$.

\begin{figure}
\begin{centering}
\vspace{-0.5cm}
\includegraphics[width=0.36\textwidth]{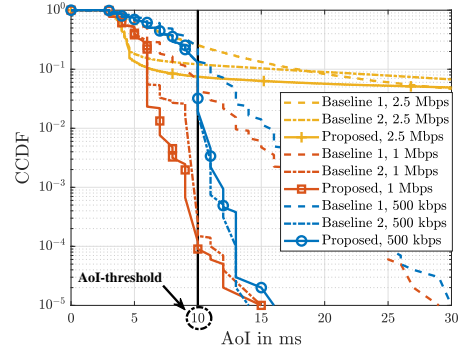}\vspace{-0.3cm}
\par\end{centering}
\centering{}\caption{CCDF of the AoI for various RB allocation schemes and different arrival
rates with $K=20$ and $M=200$.\label{fig:SchemesComparison}}
\vspace{-0.5cm}
\end{figure}
Fig.~\ref{fig:SchemesComparison} compares the AoI of the different
schemes in terms of complementary cumulative distribution function
(CCDF) for different status updates' arrival rates. Note that, due
to the learning of system dynamics $f_{k}$, the GPR-based allocation
schemes (Baseline 2 and Proposed) outperform Baseline 1 for all of
the considered arrival rates, in terms of the AoI violation probability,
i.e. $\textrm{Pr}\left\{ \triangle(t)>d\right\} $. In this regard,
the AoI violation probability is reduced by at least $52\%$, $85\%$,
and $99.6\%$, when the arrival rate is $2.5\,\text{Mbps}$, $500\,\text{kbps}$,
and $1\,\text{Mbps}$, respectively. Moreover, Fig.~\ref{fig:SchemesComparison}
shows that the AoI violation probability when the arrival rate is
$1\,\text{Mbps}$ is significantly lower than the AoI violation probability
when the arrival rate is $500\,\text{kbps}$ or $2.5\,\text{Mbps}$.
This means that the arrival rate of status updates plays a significant
role in determining the performance of the system. Furthermore, it
should be noted that exploration can further reduce the AoI violation
probability. This reduction is more pronounced when the arrival rates
are $1\,\text{Mbps}$ and $2.5\,\text{Mbps}$.

\begin{figure}
\begin{centering}
\vspace{-0.5cm}
\includegraphics[width=0.36\textwidth]{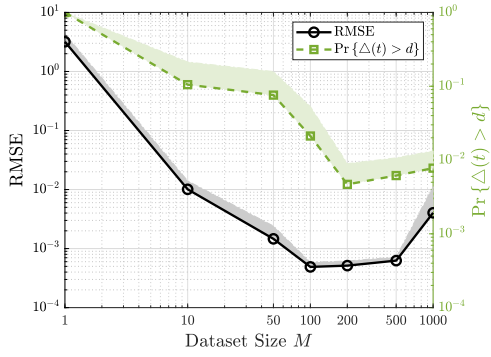}\vspace{-0.3cm}
\par\end{centering}
\caption{RMSE and AoI violation probability for different dataset sizes $M$
with $K=20$, $\alpha_{\text{c}}=1$ and $\alpha_{\text{i}}=0$.\label{fig:RMSE-Mf}}
\vspace{-0.4cm}
\end{figure}
In Fig.~\ref{fig:RMSE-Mf}, the next-slot AoI estimation performance
and the AoI violation probability are shown for the different dataset
sizes $M$ used for active learning. Here, the shaded area highlights
the upper $95\%$ confidence interval. The estimation performance
is measured using the root mean square error (RMSE) metric, which
is calculated per VUE pair $k$ as $\sqrt{\nicefrac{\sum_{t}\left(\mu_{\hat{\triangle}_{k}\left(t+1\right)}-\triangle_{k}\left(t+1\right)\right)^{2}}{T}}$.
Note that as $M$ increases, the next-slot AoI estimation becomes
more accurate, reducing the RMSE, up until $M\approx100$. However,
beyond $M=100$, RMSE increases and becomes worse at $M=1000$. This
corroborates the fact that leveraging more history about the system
is not always beneficial and may hurt the overall performance. This
is due to the fast varying vehicular environment, outdated observations
become uncorrelated, yielding poor prediction accuracy. Therefore,
the main drawback of GPR, being computationally heavy, diminishes
because accurately estimating the next-slot AoI would not require
large-sized datasets. Furthermore, Fig.~\ref{fig:RMSE-Mf} shows
that accurate AoI estimation improves decision making in terms of
minimizing the AoI violation probabilities. Following the RMSE trend,
the probability of AoI violation decreases when using up to $M=200$
data samples. The uncorrelated and outdated data samples in scenarios
with $M>200$ lead to poor prediction accuracy thereby increasing
the AoI violations probability.

\begin{figure}
\begin{centering}
\vspace{-0.1cm}
\includegraphics[width=0.36\textwidth]{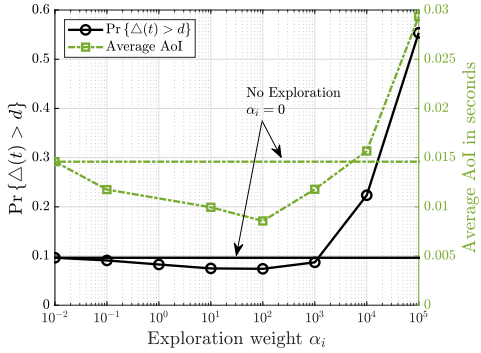}\vspace{-0.3cm}
\par\end{centering}
\caption{Exploration vs AoI violation probability and average AoI.\label{fig:ExplorationvsExploitation}}
\vspace{-0.4cm}
\end{figure}
In Fig.~\ref{fig:ExplorationvsExploitation}, we study the impact
of the exploration parameter $\alpha_{i}>0$ on the average AoI and
AoI violation probabilities. Fig.~\ref{fig:ExplorationvsExploitation}
shows that, as the exploration weight $\alpha_{i}$ increases up to
$10^{2}$, the AoI violation probability and the average AoI decrease.
This means that exploration allows each VUE pair to sample better
control decisions thus enhancing the performance. However, further
increasing $\alpha_{i}>10^{2}$ results in a performance degradation,
in terms of an increased AoI violation probability and increased average
AoI. The reason is that when the exploration weight is high, each
VUE pair will be biased towards choosing the control action that provides
more knowledge about the system instead of minimizing the AoI violation
probability. Henceforth, a balance between exploration and exploitation
needs to be taken into account.\vspace{-0.3cm}

\section{Conclusion\label{sec:Conclusion}}

In this paper, we have studied the problem of allocating transmission
power and RBs in vehicular networks under uncertainty. The main objective
is to balance a tradeoff between minimizing the probability of the
predicted AoI exceeding a certain threshold and maximizing the amount
of knowledge learned about the system dynamics in an online manner.
In this regard, GPR is used to actively learn the network dynamics
and estimate the future AoI, and a decentralized allocation approach
is proposed. Simulation results have shown a significant improvement
in terms of AoI violation probability, compared to other baselines.
Finally, our results have also shown that a balance between exploration
and exploitation is important to yield the best performance in terms
of AoI violation probability. Thus, finding the optimal dataset size
and exploration-exploitation weights to ensure a reliability target
could be an interesting topic for future work.\vspace{-0.3cm}

\bibliographystyle{IEEEtran}
\bibliography{Letter_references}

\begin{thebibliography}{10}
\providecommand{\url}[1]{#1}
\csname url@samestyle\endcsname
\providecommand{\newblock}{\relax}
\providecommand{\bibinfo}[2]{#2}
\providecommand{\BIBentrySTDinterwordspacing}{\spaceskip=0pt\relax}
\providecommand{\BIBentryALTinterwordstretchfactor}{4}
\providecommand{\BIBentryALTinterwordspacing}{\spaceskip=\fontdimen2\font plus
\BIBentryALTinterwordstretchfactor\fontdimen3\font minus
  \fontdimen4\font\relax}
\providecommand{\BIBforeignlanguage}[2]{{%
\expandafter\ifx\csname l@#1\endcsname\relax
\typeout{** WARNING: IEEEtran.bst: No hyphenation pattern has been}%
\typeout{** loaded for the language `#1'. Using the pattern for}%
\typeout{** the default language instead.}%
\else
\language=\csname l@#1\endcsname
\fi
#2}}
\providecommand{\BIBdecl}{\relax}
\BIBdecl

\bibitem{AoI_First}
S.~Kaul, M.~Gruteser, V.~Rai, and J.~Kenney, ``{Minimizing Age of Information
  in Vehicular Networks},'' in \emph{Proc.~IEEE Commun.~Soc.~Conf.~on Sensor,
  Mesh and Ad Hoc Commun.~and Netw.}, Salt Lake City, UT, USA, Jun. 2011, pp.
  350--358.

\bibitem{OurWork}
M.~K. {Abdel-Aziz}, C.~{Liu}, S.~{Samarakoon}, M.~{Bennis}, and W.~{Saad},
  ``{Ultra-Reliable Low-Latency Vehicular Networks: Taming the Age of
  Information Tail},'' in \emph{Proc. IEEE Global Commun. Conf.}, Abu Dhabi,
  UAE, Dec 2018, pp. 1--7.

\bibitem{AoIV2X_2}
\BIBentryALTinterwordspacing
A.~O. Al{-}Abbasi and V.~Aggarwal, ``{Joint Information Freshness and
  Completion Time Optimization for Vehicular Networks},'' \emph{CoRR}, vol.
  abs/1811.12924, 2018. [Online]. Available:
  \url{http://arxiv.org/abs/1811.12924}
\BIBentrySTDinterwordspacing

\bibitem{AoIV2X_3}
S.~{Zhang}, J.~{Li}, H.~{Luo}, J.~{Gao}, L.~{Zhao}, and X.~S. {Shen},
  ``{Towards Fresh and Low-Latency Content Delivery in Vehicular Networks: An
  Edge Caching Aspect},'' in \emph{2018 10th International Conference on
  Wireless Communications and Signal Processing (WCSP)}, Hangzhou, China, Oct
  2018, pp. 1--6.

\bibitem{AoIV2X_4}
Y.~{Ni}, L.~{Cai}, and Y.~{Bo}, ``{Vehicular Beacon Broadcast Scheduling Based
  on Age of Information ({AoI})},'' \emph{China Communications}, vol.~15,
  no.~7, pp. 67--76, July 2018.

\bibitem{PetarURLLC}
M.~{Angjelichinoski}, K.~F. {Trillingsgaard}, and P.~{Popovski}, ``A
  {S}tatistical {L}earning {A}pproach to {U}ltra-{R}eliable {L}ow {L}atency
  {C}ommunication,'' \emph{IEEE Trans. on Commun.}, pp. 1--1, 2019.

\bibitem{6G_WalidMehdi}
W.~Saad, M.~Bennis, and M.~Chen, ``{A Vision of 6{G} Wireless Systems:
  Applications, Trends, Technologies, and Open Research Problems},''
  \emph{{IEEE Network}}, to appear, 2019.

\bibitem{TansuetShames}
T.~{Alpcan} and I.~{Shames}, ``{An Information-Based Learning Approach to Dual
  Control},'' \emph{IEEE Trans. Neural Netw. Learn. Syst.}, vol.~26, no.~11,
  pp. 2736--2748, Nov. 2015.

\bibitem{MPC}
J.~B. {Rawlings}, ``{Tutorial Overview of Model Predictive Control},''
  \emph{IEEE Control Systems Magazine}, vol.~20, no.~3, pp. 38--52, June 2000.

\bibitem{Rasmussen2005}
C.~E. Rasmussen and C.~K. Williams, \emph{{Gaussian Processes for Machine
  Learning}}.\hskip 1em plus 0.5em minus 0.4em\relax The MIT Press, 2005.

\bibitem{Path_loss_model}
M.~Abdulla and H.~Wymeersch, ``{Fine-Grained vs. Average Reliability for {V2V}
  Communications around Intersections},'' in \emph{Proc. IEEE Global Commun.
  Conf. Workshops}, Singapore, Dec. 2017, pp. 1--5.

\bibitem{GPSignalProcessing}
F.~{Perez-Cruz}, S.~V. {Vaerenbergh}, J.~J. {Murillo-Fuentes},
  M.~{Lazaro-Gredilla}, and I.~{Santamaria}, ``{Gaussian Processes for
  Nonlinear Signal Processing: An Overview of Recent Advances},'' \emph{IEEE
  Signal Process. Mag.}, vol.~30, no.~4, pp. 40--50, Jul. 2013.

\bibitem{Stein1999}
M.~L. Stein, \emph{{Interpolation of Spatial Data}}.\hskip 1em plus 0.5em minus
  0.4em\relax Springer New York, 1999.

\bibitem{ChenEVT2018}
C.-F. Liu and M.~Bennis, ``{Ultra-Reliable and Low-Latency Vehicular
  Transmission: An Extreme Value Theory Approach},'' \emph{IEEE Commun. Lett.},
  vol.~22, no.~6, pp. 1292--1295, Jun. 2018.

\bibitem{ETSIInterVehicleDistance}
{ETSI TR 138 913}, ``{5G; Study on Scenarios and Requirements for Next
  Generation Access Technologies},'' \emph{V14.2.0}, May 2017.

\bibitem{rasmussen2010gaussian}
C.~E. Rasmussen and H.~Nickisch, ``{Gaussian processes for machine learning
  ({GPML}) toolbox},'' \emph{J. Mach. Learn. Res.}, Nov. 2010.

\end{thebibliography}

\end{document}